\documentclass[12pt]{article}
\usepackage{graphicx}
\usepackage{amssymb,amsmath,amsfonts,palatino,amsthm}
\usepackage{amssymb}
\usepackage{epstopdf}
\newcommand{\beq}{\begin{equation}}
\newcommand{\eeq}{\end{equation}}

\newcommand{\f}{\begin{equation}}
\newcommand{\ff}{\end{equation}}

\begin{document}


\title{Matrix universality of gauge and gravitational dynamics}
\author{Lee Smolin
\\
\\
Perimeter Institute for Theoretical Physics,\\
31 Caroline Street North, Waterloo, Ontario N2J 2Y5, Canada}
\date{March 18, 2008}
\maketitle

\begin{abstract}
A simple cubic matrix model is presented, which has truncations that, it is argued, lead at the classical level to a variety of  theories of gauge fields and gravity.   The latter includes  Chern-Simons theory in $d=3$, and  $BF$ theory and  general relativity in $d=4$.   General relativity coupled to Yang-mills theory for any $SU(N)$ may also arise from quantum corrections.   

On the basis of these results we conjecture that there are large universality classes of cut-off gauge and gravity theories, connected by transformations that mix up local and spacetime symmetries.  If our universe is described by one of these theories then the question of the choice of the laws of physics is to a large extent subsumed in the problem of the choice of initial conditions in cosmology.

\end{abstract}
\newpage
\tableofcontents

\section{Introduction}

It used to be widely believed that the search for the unification of the known interactions and particles within quantum theory would lead to a unique theory, knowledge of which would lead to explanations for the gauge and symmetry groups, representations and parameters of the standard model and predictions for future experiments.  Instead, string theory, the most developed approach to such a unification, appears to lead to a vast landscape of equally consistent 
theories\cite{lotc,lenny}, at least perturbatively, while non-perturbative approaches to quantum gravity also show few constraints on matter coupling\cite{me-extended}.  

There are roughly speaking two factors that may go into an explanation of why particular laws are selected from a landscape of possible laws: statistical considerations such as the Anthropic 
principle\cite{AP} and dynamical principles such as proposed in cosmological natural selection\cite{CNS,lotc}.  There are several arguments, given in detail in \cite{CNS-review}, that lead to the conclusion that statistical considerations alone cannot yield predictions that are verifiable or falsifiable.  The many recent attempts to achieve predictions from some version of statistical or anthropic considerations on the landscape have not contradicted this.  This means that any approach to a landscape of theories that leads to verifiable or falsifiable predictions must be based on a dynamical mechanism for selection of the laws that apply to our universe. 

Thus, a list of possible theories is not enough, there must be processes that allow the choice of laws to evolve as the universe does.   Thus it appears that  to  do physics on the landscape we require  a meta-theory that governs how theories evolve in time.   

But the postulation of such a meta-theory turns out to lead to still  more challenges\footnote{Note that it is not necessary to specify a meta-theory to achieve verifiable or falsifiable predictions, a statistical characterization of the dynamical mechanism as in cosmological natural selection suffices, for the same reason that Darwin and Mendal did not have to know molecular biology to have a theory of evolution that made falsifiable predictions.  Nonetheless, it is reasonable to hope that we could find a suitable metatheory to explain transitions between regions of space and time where different emergent laws hold.}.  

First, how would we verify a proposal for a meta-theory?  Suppose that two distinct meta-theories were proposed, which both allowed solutions leading to the laws we observe?  It is hard to imagine how experiment or observation in our universe could select between two proposals for meta-theories. 

Second, a meta-theory will presumably have a space of solutions, which describe different effective low energy theories.  So given a cosmological observation, it might be accounted for by the choice of meta-theory, or it might be accounted for by choice of initial conditions.  How are we to tell the difference between an observation that constrains theory choice and an observation that constrain choices of initial conditions?  

Note that this problem is already severe in cosmology. To name one example, the recent claims of an observation of 
non-gaussianity\cite{nongauss} could, depending on what paper one reads, be explained by a non-standard choice of inflationary theory or a non-standard initial condition within standard single field slow role inflation.  

It is of course possible that we have not looked hard enough for the unique consistent theory of everything, but it is also possible the search for such a theory is fruitless because it is based on metaphysical assumptions that need to  be abandoned for physics to progress.  Perhaps there simply is not a unique unified theory that also uniquely specifies the parameters of low energy physics.  

How then are we to understand why our universe appears to obey one set of laws, rather than another?

The purpose of this paper is to propose a novel approach to this question.  Suppose that a large class of theories, which included the standard model as well as a large set of plausible alternatives, were actually equivalent to each other, in the sense that there were transformations that mapped the degrees of freedom and solutions of any two of these theories into each other.  Then any solution of any of the class of theories could be mapped to any solution of any other theory of the class. 

Of course not every theory in this class would have to resemble general relativity coupled to gauge theories and chiral fermions.  It would be sufficient if only a subclass did.  

If this were the case then there would be no sense in which any of these theories could be considered more fundamental than another, nor would there be any meaning that could be given to the claim that one, rather than another, was the true theory.  The puzzle of the ambiguity of choice of initial conditions versus choice of theories would be resolved, because the only meaningful choices within the class would be choices of initial conditions\footnote{For more on the status of laws in cosmology, see \cite{inprep}.}.  Furthermore the evolution from one theory to another could be  understood in terms of quantum transitions between different semiclassical solutions of the theory.  

Before dismissing this possibility as crazy, let us take into account the various arguments  that lead to the conclusion that quantum theory plus diffeomorphism invariance forces theories to be finite, so that there are finite numbers of degrees of freedom in every quantum theory containing gravity.   In this case, each theory of gravity plus $SU(n)$ gauge fields in $d$ space-dimensions has, at least naively, roughly   
\f
N= (\frac{L}{l_p})^d (n^2 +1)
\ff
total degrees of freedom, where $L$ is the infrared cutoff given by the cosmological constant and $l_P$ is the ultraviolet cutoff given by the Planck scale.  (We neglect fermions in the following to simplify the argument.)  It seems plausible that theories with different $N$ values cannot be equivalent.  But could two gauge theories coupled to gravity be equivalent, with different dimensions and gauge groups, so long as they had the same $N$?   The demonstration of such equivalences would involve mappings between their degrees of freedom that  mix up  spacetime with internal symmetries.  That is, the transformations between theories would not respect locality.  These transformations would not be apparent from the naive continuum expressions of the theories, but they would become apparent when they were expressed in cut-off forms with finite numbers of degrees of freedom.  Were this true, the holographic principle might be a special case of a wider class of equivalences amongst theories.  

The purpose of this paper is to propose that there are indeed such  large universality class of cutoff theories of gauge fields and gravity.  This done by exhibiting a simple matrix model that has solutions and truncations which lead to a diverse set of cut-off gauge and gravitational theories, in different dimensions, with different gauge groups.  

Before presenting this theory and outlining the paper, let me mention three considerations which suggest the plausibility of this resolution of the search for a fundamental theory.

First, there are already examples of large equivalence classes amongst gauge theories of different  types. Some of the best studied of these arise in  supersymmetric gauge theories and string theory. These include conjectures of dualities between theories with different gauge groups and, as in the case of the $AdS/CFT$ conjectures, different numbers of dimensions.  Others do not require supersymmetry but involve dualities  among  non-commutative and matrix formulations of gauge theories\cite{janetal,harold03,asato}.   It is then natural to ask if all of these conjectured dualities, supersymmetric and not, may be several tips of a single iceberg involving a much wider class of dualities.  If so, the question is what principle underlies all these dualities.  

Second, consider the consequences of two widely held beliefs, that spacetime is emergent and that the theory it is emergent from is finite.  It follows that locality is also an emergent property\cite{qgraphity}.  If different spacetimes emerge by constructing different effective field theories around different solutions of the fundamental theory, then it follows that whether two degrees of freedom are related by a translation in space or by an internal symmetry transformation will not be absolute, but will depend on the solution the effective description is based on.  This makes it possible that theories with different spacetime dimensions and internal symmetries will emerge from the same fundamental dynamics.  

Third, there have been a number of suggestions that physical processes are computations\cite{phys-comp}.  However, 
the central result in computer science is the universality of computation, that all computers are equivalent to a  universal computer, a Turing machine.  Any computer can be simulated on any other computer, by writing an appropriate program.  Might it be that there is also a universality class of dynamical theories, any solution of one may be represented by a solution of another by a precise choice of initial conditions?

The metaphor of ``programming the universe", even if it is not precisely true, may give us guidance for how to proceed here.  
For, even if there is a large equivalence class of theories as described above, it may be easier to see this from one representative than another.  What is needed is something like the Turing model, a very simple representative of the class, which is  very helpful when proving the universality of computation.  For one does not have to directly demonstrate the equivalence of any two computers, one just needs to show the equivalence of each to a Turing machine.  

We then seek the equivalent of a Turing machine for gauge and gravitational theories, a simple theory from which a variety of different theories of gauge and gravitational interactions can be reproduced.  

Let us then note that there are at least three ways that a dynamical 
theory, $U$,  may give rise to another theory  $T $.    One can  plug in an ansatz to the action for $U$, leading to an action for $T$.  In this case we say that $U$ {\it truncates} to $T$\footnote{Such truncations are also called {\it constrained matrix models}\cite{harold03}.}.  Or the solutions to $U$ can include solutions to $T$, in which case we say that $U$ {\it reduces} to $T$.  They are not the same because equations of motion will be missing in a truncation that are implied by the variation of the original action-and so have to be satisfied in a reduction.  Thus, reduction is stronger than truncation.  Another possible relation  is for $T$ to arise as a low energy effective approximation to the expansion in terms of small deviations from either a truncation or a reduction of $U$.

We are then looking for a theory $U$ that has the following characteristics:

\begin{itemize}

\item{}It has a very large but finite number of degrees of freedom. 

\item{}It should truncate or reduce to cutoff versions of a large variety of different theories, including general relativity in $3+1$ dimensions, coupled to Yang-Mills theory for a variety of gauge groups, $G$.  

\item{}For reasons discussed above, the truncations or reductions will introduce notions of locality that are inconsistent with each other, as they led to theories of different dimensionality.  This suggests the theory should be truly background independent, so that the spacetime manifold on which the metric, connection and gauge fields are defined is not present when the dynamics is formulated, but emerges only from the study of special classes of solutions.

\item{}The action and equations of motion of the theory should be extremely simple, so that their physical content is minimal and the specification of kinematics and dynamics arises only by the truncation of degrees of freedom or selection of a class of solutions of the meta-theory.   

\end{itemize}

In this note I would like to propose a candidate for such a universal meta-theory and provide evidence it has truncations with the required properties.  
The degrees of freedom are
as simple as possible, they are an $N \times N$ Hermitian matrix, for a very large $N$, which will be
called $M$, with indices $M_i^j, \ \ i,i=1,...,N$.

The dynamics cannot be linear because we want its solutions to reproduce those of non-linear field equations.  The simplest non-linear dynamics are quadratic equations, which arise from a cubic action.
The simplest possible non-linear action for matrices is 
\f
S=Tr M^3
\label{action}
\ff
The theory has a symmetry under $U(N)$,  Let $U$ be an element of $U(N)$ then the action (\ref{action}) is invariant under
\f
M \rightarrow M' = UMU^{-1}
\label{symmetry}
\ff
We will see below that this gives rise to local gauge symmetries, hence (\ref{symmetry})  should be seen as a gauge symmetry.    

As we shall see below, it is not difficult to find evidence that this simple theory has reductions with the required properties.    Thus, in the next few sections we will see that this simple model has truncations that yield many of the theories of connections that physicists study.  These include topological field theories such as 
Chern-Simons theory for any $U(N)$ in three dimensions, as well as 
$BF$ theories in $4$ dimensions.  Other truncations yield theories with local degrees of freedom including general relativity in $d=4$ and massive Yang-Mills theory in $d \geq 4$.  

We also argue in section 6 that when the one loop effective action is taken into account, there are truncations, and perhaps reductions, which yield general relativity in $d=4$ coupled to Yang-Mills fields for any $U(N)$.

In section 8 we turn to the study of reductions of the full set of matrix equations of motion.   We find reductions which yield general relativity in four dimensions and a non-commutative form of Chern-Simons theory in $d=3$.  The full set of equations of motion appears to naturally include gauge fixing terms for these theories.  

It is also possible that the present proposal has a relationship to string theory.  The action (\ref{action})  also has truncations that give the bosonic sectors of  some of the cubic matrix models, studied before in \cite{cubicmatrix}, which were proposed as background independent forms of string theory.  Thus, it appears possible that at least bosonic string theories are also contained in the class of theories that arise from truncations of (\ref{action}). However the exact cubic matrix models  studied in \cite{cubicmatrix} are not obviously in the classes discussed in this paper, because they involve super Lie algebras or the exceptional Jordan algebra.  

What is certainly the case is that the idea of basing a unification on a cubic action, used in \cite{cubicmatrix} and here, was partly inspired by developments in string theory, where a background dependent field theory of open strings with an action analogous to a Chern-Simons theory\cite{open-sft} was understood to be derivable from expanding around a solution of a background independent closed string field theory whose action is a certain trace of products of string fields\cite{andy}

One may ask how all these theories can arise from a simple cubic action.  The answer is that when expressed in certain first order forms, where auxiliary fields are  used to write the actions so that only a single derivative appears, these theories all have cubic actions.  For general relativity this requires writing the theory in connection variables, such as those given by Ashtekar\cite{ashtekar} and Plebanski\cite{Plebanski}.  This is a remarkable fact, whose significance for the project of unification has perhaps been insufficiently appreciated.   There could not be a simpler form as the equations of motion are then all quadratic equations; any simpler theory would be linear.  

At a non-perturbative level the simple first order form of the actions make possible clean paths to quantization in which the hamiltonian formulations are all polynomial and the path integral measures are determined by group theory. Thus, the unification of all the above theories in a single, simple matrix model is another piece of evidence that these connection formulations of general relativity are more fundamental that the original metric formulation and are a necessary route to their quantization.  

Once it is realized that these different theories all have cubic actions the idea of unifying them by writing them as truncations of a cubic matrix model naturally suggests itself.

One can then ask how such different theories may arise from a single matrix model.  The answer, as we will see below, is that the different truncations involve different tensor product decompositions of the space of matrices.  This is analogous to the way in which emergent degrees of freedom and associated conservation laws may arise from symmetries associated with tensor product decompositions, giving rise to noiseless subsystems in  quantum mechanics\cite{nfs}. This has been proposed as the origin of the physical degrees of freedom of background independent theories in\cite{nfs-gauge,fot-dav}. 

 Finally, we may remark that there are four independent lines of argument that matrix models may underlie quantum mechanics as their ordinary statistical mechanics appears to naturally describe give a non-local hidden variables theory which approximates quantum 
mechanics\cite{adler,artem,me-qm,mefotini-qm}. 

Let me stress that the arguments in this note are incomplete and far from rigorous.  
Also, all the results discussed here are at the classical level, whether they extend to the quantum theory has yet to be investigated.  There are obvious issues quantizing a cubic action such as (\ref{action}) directly.  Nonetheless, the results found here may motivate us to try to define quantum theories corresponding, if not to the full theory, to the various truncations.  Or,   as discussed in \cite{adler,artem,me-qm,mefotini-qm} the theory  may already be quantized, in the sense that its statistical mechanics reproduces quantum mechanics in certain approximations.    

The results found so far do suggest several conjectures which, if true, would provide a realization of the picture discussed above.  These conjectures are discussed in the concluding section 9.  

\section{Compactification}

Let's start by reviewing a basic trick of matrix models called {\it compactification} which is how fields on manifolds emerge from a dynamics of matrix elements\cite{compactification,cubicmatrix}.  For simplicity we discuss here compactifications that lead to fields on tori, other compactifications leading to spheres (non-commutative for finite $N$) or other spaces also exist as 
discussed in \cite{harold03,harold2}. 

We consider that the matrices $M_{ij}$ are a truncation of operators describing the quantum mechancis of a single particle on a circle.  That is let ${\cal H}$ be the Hilbert space of quantum mechanics
on a circle and let ${\cal H}_N$ be its $N$ dimensional subspace spanned by a basis of the first $N$ fourier modes.  Let $\hat{M}$ be an operator in ${\cal H}$, then
\f
M_{nm}= <n| \hat{M}|m> =  \oint d\theta e^{i n \theta} \hat{M} e^{-i m\theta}
\ff
with $m,n < N$ is an operator in ${\cal H}_N$.  
The derivative operator is given by 
\f
 \imath [\partial_\theta ] _{nm}= <n| \partial_\theta |m> =  m \delta_{nm}
\ff
The plane wave is given by
\f
[e^{-\imath p \theta } ] _{nm}= <n| e^{-\imath p \theta }  |m> =  \delta_{n-m,\ p}
\ff
One can check that as matrices
\f
\left [ [\imath  \partial_\theta ], [e^{-\imath p \theta } ] \right ] = p [ e^{-\imath p \theta } ]
\ff

Let us consider a function $f(\theta )$ on $S^1$ with fourier expansion
\f
f(\theta )= \sum_p f_p e^{\imath p \theta}
\ff
Then its matrix representation is 
\f
f_{nm}= <n| \hat{f}|m> = \sum_p f_p \delta_{n-m-p , \  0} 
\ff
The idea in the next sections is to introduce space by means of these matrix derivatives.

Consider the commutative of two matrices, $[A,B ]$
We can consider these as operators acting on ${\cal H}_N$ and in that sense write an ansatz
\f
A= [\imath \partial_\theta ] +[ a (\theta ) ], \
\label{ansatz1}
\ff
We also consider the ansatz
\f
 B =  [ b(\theta ) ] 
\ff
Then we have
\f
[A,B] = [\imath \partial_\theta B ] + [a, B]
\ff
We see that it is natural to introduce covariant derivatives, and hence connections along with the interpretation of matrices as giving matrix elements for operators for quantum mechanics on a line. 

The ansatz (\ref{ansatz1}) corresponds to a reduction of degrees of freedom, as a general operator in quantum mechanics on the circle contains arbitrary powers of $\partial_\theta$.  Thus (\ref{ansatz1}) should be seen as the two leading terms in an expansion in powers of momentum operators of a general operator, $\hat{O}$,
\f
\hat{O}= [ a_0 (\theta ) ]+  a_1 (\theta ) [\partial_\theta ] + 
a_2  (\theta ) [\partial_\theta^2 ]  + ... + a_n  (\theta ) [\partial_\theta^n ] + ...
\label{ansatz2}
\ff

 Thus, the notion of covariant derivatives emerges to leading order in the expansion in derivative operators, which is natural in a sector of solutions in which ``space" has emerged.

Under this correspondence, the trace becomes a sum over momentum modes, which becomes in the limit $N \rightarrow \infty $ an integral over a torus.
\f
Tr M  \rightarrow \sum_p m_p \rightarrow  \oint d\theta m (\theta ) 
\ff

The point is that fundamentally,  we have a theory whose degrees of freedom are the elements of a very large matrix.  We can interpret some solutions in terms of fields living in tori, because we can make these 
ansatz's in which the matrixes are of the forms that arise from the quantum mechanics on the circle. We then additionally make an expansion in powers of derivatives and keep only the leading terms.    If these have solutions we have established, up to subtelties concerning the $N \rightarrow \infty$ limit that these ansatz's of the matrix models lead to solutions.  This is the sense in which manifolds and fields on them can arise as approximations to theories whose degrees of freedom are just elements of large matrices.

It should be stressed that what one gets naturally to leading order in the derivative expansion is the emergence of theories of connections on manifolds.  If this is done in such a way that the theory that emerges on the manifolds is diffeomoprhism invariant, then what one is going to get naturally is diffeomorphism invariant gauge theories.  The simplest of these, as we will see in the next two sections,  are topological field theories.  The next simplest are theories constructed as perturbations or deformations of topological field theories, which are general relativity and Yang-Mills theories.  

In the following we will just keep terms to zeroth and first order in the expansion of matrices around solutions from which space emerges.  The theories we will find are then all just leading order effective descriptions of the low energy behavior of the fundamental matrix model. Finally, we should also stress that if $N$ is large, but not infinite, then the correspondence to fields on manifolds is always approximate.  

\section{Matrix Chern-Simons theory}

To motivate the truncations we study we may recall the idea of noiseless or decoherence free subspaces, from quantum information theory. In these systems,  persistent, physical degrees of freedom are brought into existence by splitting the system into subspaces, in a way that introduces
symmetries in the interactions of those subspaces with their environments\cite{nfs}.  Indeed it has been argued that in some cases emergent gauge symmetries can be understood as arising from noiseless subsystems\cite{nfs-gauge}, and that this may be the origin of physical degrees of freedom in background independent systems\cite{fot-dav}.  While our context is different because we are studying classical dynamics of matrices rather than quantum mechanics, we can here also study the effects of truncations based on tensor product decompositions.  A simple one we may begin with is
\f
M= \sum_{a=1}^3 A_a \otimes \tau^a
\label{ansatz22}
\ff
where $\tau^a$ are the three $2 \times 2$ Pauli matrices and $A_a$ are three $M \times M$ matrixes, where $M=N/2$.  Then the action becomes 
\f
S= Tr (A_a A_b A_c )  Tr(  \tau^a \tau^b \tau^c ) =  Tr (A_a A_b A_c )  \epsilon^{abc}
\ff
We can go further.  Let $T^I $ be the $n$ generators of a Lie algebra, $G$ in an $r$ dimensional representation.  Then we write
\f
M=  A_{a I} \otimes T^I \otimes  \tau^a 
\ff
where now $A_{aI}$ are $3n$ matrices each $M \times M$, where now $N= 2rM$.

The action is now,
\f
S= Tr (A_{aI} A_{bJ} A_{cK} )  Tr(  \tau^a \tau^b \tau^c ) Tr( T^I T^J T^K )=  Tr (A_{aI} A_{bJ} A_{cK} )  \epsilon^{abc} f^{IJK}
\ff

We now apply the compactification trick discussed in the last section three times,  to each of the  three matrixes $A_{a}$, to bring into being three circles.  We write the $M\times M$ matrixes $[A_{a}]_i^j$ as follows.  We divide the indices
$i,j= 1,...M$, into a product of two indices $ \bar{i}=1,...,P$ and $\alpha = 1,..., r$ so that $M=rP$.
Thus, $i = \bar{i}\alpha $ and 
\f
[A_{a}]_i^j= [A_{a}]_{i\alpha} ^{j \beta}= [A_{aI}]_{i} ^{j }T_\alpha^{I \beta}
\ff 
Now we use the compactification trick three times to write
\f
[A_{a}]_{\bar{i} \alpha}^{\bar{j} \beta} = [\partial_a]_{\bar{i}}^{\bar{j}} \delta_\alpha^\beta
+[a_{a} (x,y,z) ]_\alpha^\beta 
\label{trick1}
\ff
We note that we can pick the derivative operators so that 
\f
[\partial_a , \partial_b ] =0
\ff

The trace, which in the ansatz is a sum over momentum modes, gives rise,  in the limit $N \rightarrow \infty$, to an integral over the three torus
\f
Tr \rightarrow \int_{T^3} Tr_r
\ff
where $Tr_r$ is trace over the $r$ dimensional matrixes.  
So that we have
\f
S= \int_{T^3} d^3x \epsilon^{abc} Tr_r  \left ( a_a \partial_b a_c +\frac{2}{3} a_a a_b a_c 
\right )
\ff

It is interesting to note that the original meta-action, (\ref{action}) has a global symmetry, which is
$U(N)$.  This has, under the ansatzes (\ref{ansatz2},\ref{trick1}) this has become a local gauge invariance, under an arbitrary gauge group, $G$, on a three manifold, $T^3$.  
\f
\delta A_a = {\cal D}_a \lambda = \partial_a \lambda +[A_a, \lambda]
\label{localgauge}
\ff
We see that we have realized the idea of the emergence of gauge symmetries from noiseless subsystems, proposed in (\cite{nfs-gauge}).

\section{BF Theory}

Now that we have the basics we add some complexity which will move us up one dimension.  Let $\gamma^a$ be the four Dirac gamma
matrixes-from now on $a,b,c= 0,1,2,3$ and let us write $\gamma^{ab}= [\gamma^a , \gamma^b]$. 
We choose an ansatz for solutions of the theory defined by (\ref{action}),
\f
M=  A_{a I} \otimes T^I \otimes \gamma^a  + B_{ab I} \otimes T^I \otimes \gamma^{ab}\gamma_5
\ff
The action (\ref{action}) now is
\f
S = Tr (A_{aI} A_{bJ} B_{cdK} )  \epsilon^{abcd} f^{IJK}
\ff
Using the compactification trick we  now have the emergence of a four-torus,
\f
Tr \rightarrow \int_{T^4} Tr_r
\ff
so the meta-action yields $BF$ theory\cite{BF}
\f
S = \int_{T^4} d^4x \epsilon^{abcd} B_{cdK} F_{ab}^K    
\ff

\section{The  Plebanski action}

To get the Plebanski action we need to add one degree of freedom, which are the $G$ matrixes of scalar
fields $\Phi_{IJ}$ and specify $G=SO(3,1)$.  We do this by expanding the ansatz to 

\f
M=   \Phi  \otimes \gamma_5 + A_{a I} \otimes T^I \otimes \gamma^a  + B_{ab I} \otimes T^I \otimes \gamma^{ab}\gamma_5
\label{ansatzP}
\ff

The action (\ref{action}) now is
\f
S = Tr (A_{aI} A_{bJ} B_{cdK} )  \epsilon^{abcd} f^{IJK} 
+ Tr (\Phi_{IJ} B_{abI} B_{cdJ} )  \epsilon^{abcd}  
\ff
where
\f
\phi^{IJ}_{mn}= [ \phi_{mn}]_i^j (T^I T^J)^i_j
\ff
We will also impose the constraint
\f
 [ \phi_{mn}]_i^i = -\Lambda \delta_{mn}
\label{cosmo}
\ff
where $\Lambda$ is the cosmological constant.

When we again do the compactification trick we find the  Plebanski action
\f
S = \int_{T^4} d^4x \epsilon^{abcd} \left [  B_{cdK} F_{ab}^K    
 +  \Phi_{IJ} B_{ab}^I B_{cd}^J  
\right ] 
\label{Pl}
\ff

Note that without the constraint (\ref{cosmo}) the  theory we have has one less field equation than general relativity, it is GR without a hamiltonian constraint imposed.

\section{Extending Plebanski}

It is interesting to see how the Plebanski action is a truncation of our simple cubic matrix model, but it would be nicer if one did not have to impose the constraint (\ref{cosmo}).  There is, to see this 
  recall from the work of Krasnov\cite{kirill} that on the quantization of the  action (\ref{Pl}) there  arise terms in the effective action of the form 
\begin{eqnarray}
S^\hbar  &=& \int_{T^4} d^4x \epsilon^{abcd}  f (\Phi_{IJ} \Phi^{IJ} ) (B_{abK} B_{cd}^K ) 
\nonumber \\
&=& \int_{T^4} d^4x \epsilon^{abcd}  \Phi_{IJ} \Phi^{IJ}  (B_{abK} B_{cd}^K ) +...
\label{Shbar}
\end{eqnarray}

We can conjecture that the same terms arise in the expansion of the action (\ref{Pl}) {\it without the contraint (\ref{cosmo}) imposed.}  

We note that so long as $N$ remains finite these are finite terms in the low energy effective action, there need not be terms in the fundamental action corresponding to them. This is reasonable, we may take, for example, $N = (\frac{L}{l_{P}})^4$ as an estimate for the sizes of the matrixes we need to describe a world with cosmological constant $\Lambda = \frac{1}{L^2}$.   We then get a theory which is to leading order general relativity, but with no need to impose the constraint (\ref{cosmo}).    Thus, we see that the effective action does for large but finite $N$ reproduce general relativity.  

To add Yang-Mills fields we follow the route of \cite{me-extended}, which is to extend  $G$  to a  Lie algebra that contains
the Lorentz algebra as a proper subalgebra, keeping  the ansatz (\ref{ansatzP}).  
It is natural to hypothesize that when this is done there continue to arise terms in the effective action 
of the form (\ref{Shbar}).  If so, then the effective action  is of the form,
\f
S +S^\hbar = \int_{T^4} d^4x \epsilon^{abcd} \left [ B_{cdK} F_{ab}^K    
   + \Phi_{IJ} B_{abI} B_{cdJ}  +
 \Phi_{IJ} \Phi^{IJ} B_{abI} B_{cdJ}  +...
\right ] 
\ff

This is the extended Plebanski action that, as shown in 
\cite{me-extended}, gives rise to general relativity coupled to Yang-Mills theory for an gauge group $G$ that is contained in $G/SO(3,1)$.  Hence, if the hypothesis we made  above is true, the meta-action
(\ref{action}) has ansatzes that give rise to general relativity coupled to Yang-Mills theory with arbitrary gauge groups.  

\section{Massive Yang-Mills theory}

There are also truncations of the basic action (\ref{action}) that yield Yang-Mills theory, but with a mass, for any $U(N)$ for any $d \geq 4$.  Let us exhibit the four dimensional case.
We choose the truncation (with the same conventions as the last two sections)
\f
M=I + A_a \otimes \gamma^a - B_{ab} \otimes \gamma^{ab}
\label{ansatz-ym}
\ff
and find easily that
\f
S=TrM^3 = Tr_r \left (  -[A_a, A_b ]B^{ab} + B_{ab}B^{ab}  + A_a A_a \right )
\label{ym1}
\ff
This is just the matrix reduction of massive Yang-Mills for any $U(N)$, as we can see
by eliminating $B_{ab}$ using its equations of motion to find 
\f
S=-\frac{1}{4} Tr_r \left (  [A_a, A_b ][A^a, A^b ] + A_a A_a \right )
\label{ym2}
\ff

\section{Full reductions of the theory}

What we have done so far is to truncate the degrees of freedom, then vary to find equations of motion.
The results are interesting, but more interesting would be to find the equations of motion and then reduce the degrees of freedom on solutions.

\subsection{The full three dimensional sector}
Let us see how that works with the ansatz that gave the cubic matrix model and Chern-Simons theory.  We have to extend (\ref{ansatz2})  to a general ansatz, which is 
\f
M= \sum_{\mu= 0}^3 A_\mu \otimes \tau^\mu
\label{ansatz2b}
\ff
where $\mu=0,1,2,3=1,a$.  The $2 \times 2 $ matrixes are spanned by 
$\tau^{\mu=1,2,3}$ together with $\tau^{0} = Id$. 
$A_\mu$ are then  four $M \times M$ matrixes, where $M=N/2$.  

Then the action becomes 
\f
S= Tr (A_\mu A_\nu A_\sigma )  Tr(  \tau^\mu \tau^\nu \tau^\sigma ) =  Tr (A_a A_b A_c )  \epsilon^{abc}
-3 Tr (A_0 A_a A_a ) + Tr (A_0^3)
\label{cs2b}
\ff
The equations of motion are
\begin{eqnarray}
\epsilon^{abc} [A_b, A_c ] + 2  \{ A_0 ,A_a \} &=&0  ,
\label{eom2b}
 \\ 
A_0^2 &=& A_a A_a
\label{eom2c}
\end{eqnarray}

Let's use the  symmetry (\ref{symmetry}) to analyze these equations
(\ref{eom2b},\ref{eom2c}).   We can use the symmetry to diagonalize $A_0$, so that
\f
A_0 = \mbox{diag}[a_0,a_1, ... a_N ]
\label{diagonal1}
\ff
There will be solutions to (\ref{eom2b},\ref{eom2c}) for every choice of eignvalues in the list $a_i$.   But interesting things happen for solutions where there are degeneracies.  So let us assume that there are a fewer number, $P <N$  of degenerate eigenvalues $a_\alpha$ with $\alpha = 0,...,P-1$, each having a degeneracy $d_\alpha$.  

Each $N\times N$ matrix splits into blocks, the $\alpha,\beta$'th block has 
dimension $d_\alpha \times d_\beta$.  The matrix $A_0$ is block diagonal, the other 
$A_a$ divide into blocks which may be labeled $[A_a]_\alpha^\beta$.

We may look for solutions where the off diagonal blocks vanish, so
$[A_a]_\alpha^\beta =0 $ for $\alpha \neq \beta$, we may then label the
diagonal blocks $[A_a]_\alpha^\alpha \equiv A_a^\alpha$.

The equations of motion (\ref{eom2b},\ref{eom2c}) then split into $P$ decoupled
systems
\begin{eqnarray}
\epsilon^{abc} [A_b^\alpha , A_c^\alpha  ] =  a_\alpha A_a^\alpha  ,
\label{eom3b}
 \\ 
 A_a^\alpha  A_a^\alpha = a_\alpha^2  I_\alpha 
\label{eom3c}
\end{eqnarray}

These are the equations for Chern-Simons theory on the fuzzy sphere, which have been studied before before in \cite{harold03}.  
Each sector describes Chern-Simons theory on a non-commutative  manifold
whose derivatives satisfy (\ref{eom2b}),
\begin{eqnarray}
[\partial_a, \partial_b ] &=& a_\alpha  \epsilon_{ab}^c \partial_c
\label{eom4b}
\\
 \partial_a^2  &=& a_\alpha^2  I_\alpha 
 \label{eom4c}
\end{eqnarray}
So we see that the eigenvalue $a_\alpha$ is a parameter measuring the non-commutativity of the manifold.   If one expands around these solutions, following (\ref{trick1}), the equation of motion 
becomes the vanishing of the non-commutative curvature
\f
{\cal F}^{a_\alpha}_{ab}  \equiv \partial_a a_b - \partial_b a_a 
-a_\alpha  \epsilon_{ab}^c a_c +[a_a , a_b] =0
\ff
The equation (\ref{eom3c}) becomes a gauge fixing condition
\f
\partial_a a_a + a_a a_a =0
\ff
Note that if there is a sector $a_\alpha = 0$ this corresponds to ordinary 
Chern-Simons theory.  The non-commutativity may then be considered a 
regularization that allows us to define
\f
\lim_{a_\alpha \rightarrow 0} \partial_a^2  = 0
\ff

We can put back in the off-diagonal sectors $[A_a]_\alpha^\beta$.   If these matrices are sparse they will introduce interactions between the fields on the tori on which the
Chern-Simons theories are defined, similar to the mechanism discussed in \cite{cubicmatrix}.

\subsection{The full four dimensional sector}

We now consider all the degrees of freedom in the case corresponding to four
spacetime dimensions.  The full expansion of the $16$ dimensional space of
$4\times 4$ matrices yields
\f
M=\psi \otimes I + + \phi \otimes \gamma^5 +  A_a \otimes \gamma^a + C_a \otimes \gamma^a \gamma^5 + B_{ab} \otimes \gamma^{ab}\gamma^5
\label{ansatz-full}
\ff
The action is now
\begin{eqnarray}
S=TrM^3 &=&  Tr_r  \{  
\epsilon^{abcd} [ B_{ab}^I ([A_c, A_d]_I -[C_c, C_d]_I ) + \phi_{IJ} B_{ab}^I B_{cd}^J ]
 \nonumber \\
  &+& \psi^3   +3  \psi  [ \phi^2+ A_a A^a - C_a C^a +  B_{ab}B^{ab}]  
  \nonumber \\
  &+&  [A_a,C_b]^I B_I^{ab}  + \phi_{IJ} A_a^I C^{aJ }   \}
\label{action-full}
\end{eqnarray}

The corresponding equations of motion represent the full variation of the
action (\ref{action}) under the full parameterization (\ref{ansatz-full}).
\f
\epsilon^{abcd} [A_b, B_{cd}]^I +[C_b, B^{ab}]^I +\phi^I_J C^{aJ}
- (\psi A^a )^I = 0
\ff
\f
[A_a, A_b ]^I - [C_a, C_b ]^I = 2 \phi^I_J B_{ab}^J +
2 \epsilon_{abcd} (\psi B^{cd})^I +[A_a, C_b]^I
\ff
\f
\epsilon^{abcd}B^I_{ab} B^J_{cd} + A_a^I C^{aJ} +\{ \psi, \phi  \}^{IJ}=0
\ff
\f
[A_b , B^{ab}]^I +\phi^I_J A^{aJ} -2(\psi C^a )^I - \epsilon^{abcd}[C_b, B_{cd}]^I =0
\ff
\f
-\psi^2= \phi^2 +A_aA^a - C_a C^a + B_{ab}  B^{ab}
\ff
where $X^I=Tr (X T^I)$. 
We can now {\it reduce} the theory by setting 
\f
\psi = C_a =0
\label{reduce1}
\ff  
This results in equations of motion 

\f
\epsilon^{abcd} [A_b, B_{cd}]^I  = 0
\label{g1}
\ff
\f
[A_a, A_b ]^I = 2 \phi^I_J B_{ab}^J 
\label{g2}
\ff
\f
\epsilon^{abcd}B^I_{ab} B^J_{cd} =0
\label{g3}
\ff
\f
[A_b , B^{ab}]^I +\phi^I_J A^{aJ} =0
\label{g4}
\ff
\f
 \phi^2 +A_aA^a + B_{ab} B^{ab} =0
\label{g5}
\ff

The first three of these, (\ref{g1}-\ref{g3}) are the equations we had before, for an extended Plebanski theory.   The last (\ref{g5}) is very similar to the equation of the 
full three dimensional theory that became the gauge fixing for the internal Yang-Mills invariance on compactication, and we can conjecture it plays the same role here.  It is then natural to also conjecture that (\ref{g4}) gives on compactification a gauge
fixing of the four dimensional diffeomorphism invariance.  

Thus, it is plausible that further development which includes the one loop effective action will  lead to the conclusion that general relativity in $d=4$ coupled to Yang-Mills fields for any $U(N)$ is a reduction as well as a truncation of the cubic matrix model given by
(\ref{action}).  

Is the same the case for Yang-Mills theory in four dimensions.  To investigate this we
consider a different reduction given by 
\f
\phi = C_a =0
\label{reduce2}
\ff
This gives a rather different set of reduced equations of motion, 
\f
\epsilon^{abcd} [A_b, B_{cd}]^I - \psi A^a = 0
\label{y1}
\ff
\f
[A_a, A_b ]^I = 2\psi \epsilon_{abcd} B^{cd}
\label{y2}
\ff
\f
\epsilon^{abcd}B^I_{ab} B^J_{cd} =0
\label{y3}
\ff
\f
[A_b , B^{ab}]^I   =0
\label{y4}
\ff
\f
-\psi^2= A_aA^a + B_{ab} B^{ab}
\label{y5}
\ff

We proceed as above by diagonalizing $\psi$ and studying degenerate sectors
labeled by eigenvalues $\lambda_\alpha$.  Setting to zero off diagonal matrix 
elements in the blocking defined by these sectors, we get as above decoupled equations
for every degenerate eigenvalue $\lambda_\alpha$.  

(\ref{y2}) then becomes in each sector 
\f
 B^{ab}=\frac{1}{2\lambda_\alpha} \epsilon^{abcd}[A_c, A_d ]^I
\label{yy2}
\ff
which means (\ref{y4}) is solved identically.  (\ref{y1}) is {\it almost} the Yang-Mills
equation
\f
 [A_b, [A^a, A^b] ]- \lambda_\alpha A^a = 0
\label{yy1}
\ff
It does become the Yang-Mills equation if we take a limit 
$\lambda_\alpha \rightarrow 0$.   The last equation (\ref{y5}) can again 
be  interpreted as a gauge fixing equation in each sector
\f
-\lambda_\alpha^2= A_aA^a + B_{ab} B^{ab}
\label{y6}
\ff
If we consider a limit in which $\lambda_\alpha \rightarrow 0$ and compactify
we then find the usual Yang-Mills equations together with a gauge fixing condition
\f
\partial_a a^a + a_a a^a = - F_{ab}F^{ab}
\ff

The problem is that there is one more equation (\ref{y3}), which now reads
\f
\epsilon^{abcd} [A_a, A_b]^{(I}  [A_c , A_d ]^{J)} =0
\label{yy3}
\ff
This leads to a strong reduction of the solutions to the Yang-Mills equations.  
In contrast to the previous result, extending from the truncation which gives the
four dimensional Yang-Mills theory to the full four dimensional parameterization of the theory results in a severe reduction of the full Yang-Mils theory 

\section{Conclusions and a conjecture}

There are several lines of investigation which are needed to examine the implications of these results and constructions.   First, the quantization needs to be studied.  There is no obstacle to introducing one time dimension and then proceeding to construct the hamiltonian or path integral quantum theory based on it.  However, given that the action is not bounded from below, there are good reasons to doubt that a sensible quantum theory exists for the whole of (\ref{action}). Perhaps quantum theories are only defined for truncations or in the neighborhoods of solutions which define an emergent spacetime.  The alternative mentioned above, that quantum mechanics arises from matrix models, should also be further investigated.  

One obvious part of the natural world that is left out is fermions.  One can of course simply extend the model by inventing a fermionic matrix, $\Psi$ and adding to the action a term
\f
S^\Psi = Tr \bar{\Psi} M \Psi
\ff

Alternatively one can investigate the possibility that fermions emerge due to one of several proposed 
mechanisms\cite{me-extended,fermions}.  

Returning to the theme of the introduction, these results suggests a speculative conjecture concerning dualities amongst a large class of theories.

\begin{quotation}

{\bf Conjecture:} There is a universality class of cutoff background independent theories of connections, which are all equivalent to each other in the sense that there are  transformations that map the degrees of freedom of any one into the degrees of freedom of the other,  preserving the equations of motion.  Representatives of this equivalence class include general relativity in $d=4$ coupled to Yang-Mills theory for any $U(N)$ (plus certain degrees of freedom required to complete the unification) as well as
Chern-Simons theory, matrix Chern-simons theory, as well as the matrix model 
(\ref{action}).  

\end{quotation}

We note that if the conjecture is true, within this universality class the equivalence relations mix up local spacetime and internal symmetry structures. That is, degrees of freedom that are at the same point, or the same momentum mode in one theory are mapped to degrees of freedom at different sites or momentum modes in another theory.  Whether two matrix elements correspond to degrees of freedom at the same point or at different sites related by spacetime translations, and whether they are related by an internal or gauge symmetry or not, depends on the background the theory given by (\ref{action}) is expanded around.   We may then 
hypothesize that our universe is described by laws arising from solutions of a universal theory  of the kind described here.  

If true, this hypothesis imply that the question of why particular laws and degrees of freedom are observed in our universe is just part of the question of what chooses the initial conditions in cosmology.    A universality of dynamics such as conjectured  here would mean that the question of what is the "ultimate theory" may play as minor a role in physics as the choice of physical computer plays in computer science.

\section*{ACKNOWLEDGEMENTS}

 I am very grateful to Harold Steinacker for informing me of his prior work on constrained matrix models in \cite{harold03,harold2} and for comments on a draft of this manuscript.  I am also very grateful to Jan Ambjorn, Kostas Anagastopolous,  Laurent Freidel, Jaron Lanier,  Michael Neilsen and Edward Witten for extremely helpful comments. I also thank Asato Tsuchiya for kindly informing me of related results which I was unfortunately unaware of, which appeared previously in \cite{asato}. 
 
These ideas emerged from a project on the issues of time and the nature of law in cosmology with Roberto Mangabeira Unger and I am grateful to him for his continual stimulation and provocation.  Research at Perimeter Institute is supported in part by the Government of Canada through NSERC and by the Province of Ontario through MEDT.

\end{document}